# Contact Force Mediated Rapid Deposition of Colloidal Microspheres Flowing Over Microstructured Barriers

*P. Prakash[1,a),*], A. Z. Abdulla[2,b)], M. Varma[1,3,*]*

Author to whom correspondence should be addressed: *pp467@cam.ac.uk or *mvarma@iisc.ac.in

[1]Centre for Nanoscience and Engineering, Indian Institute of Science, Bangalore, 560012, India
[2]Department of Physics, Indian Institute of Science, Bangalore, 560012, India
[3]Robert Bosch Centre for Cyber Physical Systems, Indian Institute of Science, Bangalore, 560012, India.
a) Current affiliation: Department of Applied Mathematics and Theoretical Physics, University of Cambridge, Wilberforce Road, Cambridge, CB3 0WA, United Kingdom.
b) Current affiliation: Laboratoire de Biologie et Modelisation de la Cellule, ENS de Lyon, 69364 Lyon Cedex 07, France.

## ABSTRACT

Deposition of particles while flowing past constrictions is a ubiquitous phenomenon observed in diverse systems. Some common examples are jamming of salt crystals near the orifice of saltshakers, clogging of filter systems, gridlock in vehicular traffic etc. Our work investigates the deposition events of colloidal microspheres flowing over microstructured barriers in microfluidic devices. The interplay of DLVO, contact and hydrodynamic forces in facilitating rapid deposition of microspheres is discussed. Noticeably, a decrease in the electrostatic repulsion among microspheres leads to linear chain formations, whereas an increase in roughness results in rapid deposition.

## INTRODUCTION

A deposition process is a gradual build-up of heaps of cellular or particle milieu over time. It usually entails simultaneous action of DLVO, contact and hydrodynamic forces[1]. The dominant physical force promoting deposition is system dependent. For example, the aggregation of colloidal particles is driven by DLVO forces[2], jamming events are supported by contact force[3], and active crystals can be formed entirely due to hydrodynamic interactions[4,5]. Besides, there are other frequently ignored phenomena such as many-body interaction, tribological effects etc.[6–8] The understanding of deposition processes is also highly relevant for various industrial processes. Several technologies rely on suppression of clogging in porous channels such as membrane filters, deep-bed filtration technology, inkjet printing, porous catalyst pellets etc.

Competing physical phenomena act at once, making it exceptionally challenging to pin down the dominant mechanism during deposition events. Microfluidic devices have long been used to decipher several aspects of the deposition process. The channel geometry is one of the most explored aspects of clogging with numerous studies on the role of connectivity, tortuosity, constrictions and porosity[9–15]. Several numerical studies as well have guided the experiments by proposing clogging phase diagrams[16,17]. The simplest form of clogging occurs during pore blockage from particles whose size is larger than the pore size[18]. However, clogging is common, even when flowing particles are much smaller than the pore size. At low volume fractions (~1% w/v), the collective effects are negligible and clogging proceeds as a single particle deposition process, whereas at high volume fraction (~20% w/v) clogging is influenced by the collective behaviour of many interacting particles[19–21]. The three major physical interactions in a deposition process are particle-fluid, particle-particle and particle-surface, typically tuned by varying flow rate and ionic strength[22–25].

There is very little control on the spatial location where the clogging occurs in a microfluidic channel, making it tricky to study at a single particle level. We bypass this problem by fabricating microstructures over glass substrates allowing precise deposition of the first particle[26]. The deposition process is then explored using the first particle as a seed, from a single particle to a large accumulation. Interestingly, colloidal microspheres form a linear chain at low ionic strength whereas, a slight increase in the roughness of microspheres leads to rapid deposition. The role of roughness in the deposition process is not understood and so far is explored only in the context of shear-thickening of colloidal suspensions[27]. Our work lays out the physical and chemical condition required for rapid deposition of colloidal microspheres giving valuable insights into clogging phenomena. Further, the size of microspheres (10 µm + 4 µm) used is very similar to the size of cellular milieu (Neutrophils: $12-14$ µm, Platelets: $2-3$ µm, RBCs: $6-8$ µm) in arteries[28]. Hence, the deposition events of colloidal microspheres are reminiscent of arterial plaque formation caused by the gradual build-up of cells in arteries leading to heart attack[29–31].

## MATERIALS AND METHODS

The deposition of polystyrene microspheres of size 10 µm (density = 1.05 gm/mL) is studied by flowing them over wedge-like fabricated structures in a microfluidic device

(Figure 1(a)). The wedge microstructures are fabricated by patterning 2.2 μm thick and 5 μm wide photoresist (S1813) in an array over glass substrates. Subsequently, resist patterned substrate is isotropically etched using HF based etchant (10% HF + 7% HCl) for 15 sec. This process yields wedge-shaped symmetric microstructures of height ∼1 μm and width ∼6 μm (Figure 1(b))[26]. A PDMS (Polydimethylsiloxane) microfluidic device of height 100 μm and width 500 μm is then clamped on the microstructured substrate to seal the channel. The device is imaged at 30 fps by a custom-built inverted CMOS camera (Thorlabs) assembly illuminated using ring-shaped light (Supporting Information (SI) Figure 1(a)). Four repeats were carried out in each case to study the deposition trends across different scenarios. Before each repetition, the substrate was washed with the piranha solution ($3H_2SO_4 + H_2O_2$) to maintain its pristine condition.

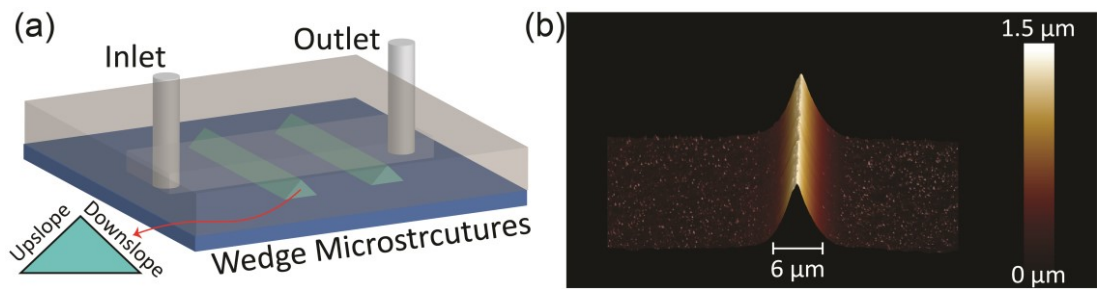

**Figure 1.** (a) Schematic of a PDMS microfluidic device (h = 100 μm, w = 500 μm) affixed over wedge microstructures. Flow is driven from Inlet to Outlet first encountering upslope and then downslope. (b) AFM of microstructures (h = 1.5 μm, w = 6 μm).

## RESULTS AND DISCUSSION

**Pinning of Microspheres in DI Water Medium.** The colloidal microspheres of size 10 μm (0.5 % w/v) in de-ionised ultra-filtered water (DI water) selectively pin on the downslope (Figure 1(a)) side of microstructures when passed at a flow rate of $10 - 50$ μl/min (Figure 2(a), SI Video 1). For lower flow rates (i.e. < 10 μl/min), microspheres also pin on the upslope in addition to downslope due to their inability in crossing upslope gravitational potential barrier (SI Video 2). The 10 μl/min is the lowest flow rate with no upslope pinning[26]. For higher flow rates we still do not see upslope pinning, however, rate of pinning drops drastically (SI Figure 1(b)). Due to the high pinning rate and precise downslope pinning, we initiated all the subsequent experiments at a flow rate of 10 μl/min.

The horizontal drag force on microspheres at a flow rate of 10 μl/min is 0.12 nN regardless of pinning in the upslope or downslope region as estimated from the hydrodynamic simulation (SI Section 2) performed using 'COMSOL Multiphysics'[32]. However, microspheres experience a lift force (11.7 pN) on the upslope instead of a downward thrust (−13.5 pN) on the downslope, explaining preferential pinning on the downslope. Interestingly, microspheres do not pin beyond a flow rate of 50 μl/min (drag force ~0.6 nN), however, once pinned (i.e. at lower flow rates ≤ 50 μl/min), they do not detach even at a very high flow rate of 300 μl/min experiencing a shear drag force of ~4.9 nN. This suggests that over time microspheres settle to a position of higher local stability. It is possible that the microspheres with a rough surface realign, leading to increased frictional contact area and adhesion. The microspheres remain pinned even if the device is flipped upside down suggesting the role of attractive DLVO force as well. Below, we have explored the deposition rate by introducing a salt solution to enhance the attractive DLVO force, and microspheres of higher roughness to increase the frictional force.

**Linear Chain Formation in the Presence of Salt Medium.** Microspheres at a flow rate of 10 μl/min and in DI water medium, selectively pin on the downslope[26]. The same 10 μm microspheres (0.5 % w/v), when flown in 0.1 M phosphate buffer saline (PBS), form linear chains (SI Video 3, Figure 2(b)). At higher PBS concentration (1 M), the colloidal solution is unstable and readily formed aggregates whereas, at a lower concentration (0.01 M), the behaviour is similar to single pinning events seen in DI water medium. The chain formation in 0.1 M PBS medium proceeds by pinning of single microspheres on the downslope, which then acts as a seed for successive microspheres to deposit one by one growing into a linear chain (SI Video 3). The microstructure facilitates pinning of only the first particle, from thereon the dynamics of chain formation is governed by particle-particle interaction. Close contact between incoming and already pinned microspheres is crucial for the microspheres to glide their way to form linear chains.

We define the fractional area coverage as the ratio of space occupied by microspheres to the total space on the substrate (Figure 2(c)). The space occupied by a microsphere is assumed to be a square of side length equivalent to the microsphere diameter as it becomes unavailable for the next incoming microsphere. The fractional area coverage in DI water (inset Figure 2(c)) saturates over time due to the filling of all available pinning sites, i.e. microstructures. In the presence of PBS (Figure 2(c)), microspheres grow into linear chains (SI Video 3), thereby propelling the area coverage up to 70% (SI Figure 4(a)). The microsphere flown in DI water pins only on microstructures, and hence the area of coverage saturates quickly, whereas it takes longer for the area coverage to saturate in PBS (0.1 M) due to the growth of microspheres into linear chains.

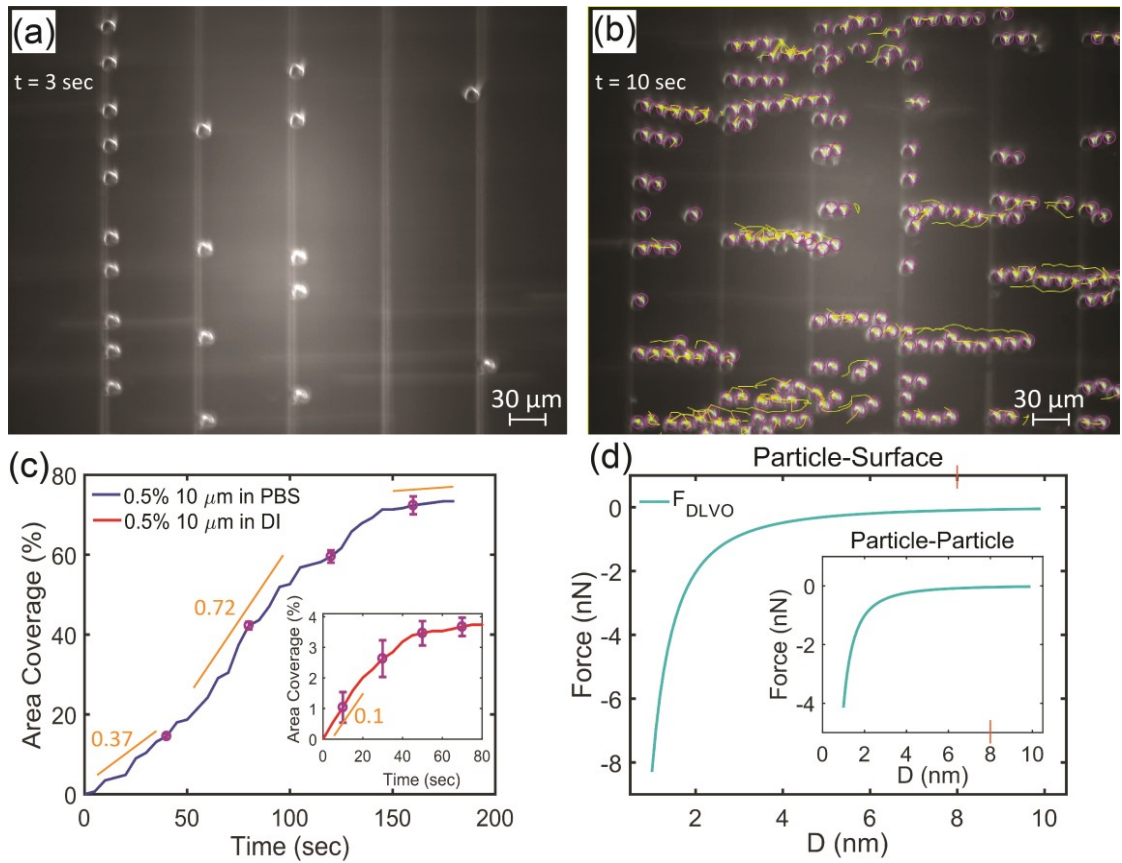

**Figure 2.** Deposition of 10 μm microspheres at a flow rate of 10 μl min$^{-1}$. (a) Microspheres in the presence of DI water pin on the downslope of wedge microstructures. (b) Microspheres in the presence of 0.1 M PBS deposit in the form of linear chains. (c) The fractional area coverage of microspheres flowing in 0.1 M PBS vs time. The curve saturates after ∼2 min. Inset figure shows area coverage when microspheres pin only on the downslope in the presence of DI water. (d) DLVO force between microsphere of size 10 μm and glass substrate in DI water medium. Inset image shows DLVO force between two microspheres of size 10 μm in DI water medium.

The interplay among hydrodynamics, DLVO ($F_{DLVO}$) and contact forces are responsible for this fascinating linear alignment of microspheres. The DLVO force comprises of electrostatic double-layer repulsion ($F_{EDL}$) and attractive van der Waals force ($F_{VDW}$). The electrostatic potential of microspheres determined by zeta potential measurement drops from $-31.5 \pm 1.7$ mV in DI water to $-4.33 \pm 0.86$ mV in 0.1 M PBS. The corresponding Debye length ($\sim 0.3/\sqrt{[C]}$ nm, C – buffer concentration in M) drops from 948.7 nm in DI water to ~1 nm respectively[33]. The decrease in electrostatic potential and Debye length in PBS medium leads to lower repulsion ($F_{EDL}$) and therefore increases the probability of contact, allowing microspheres to graze closely resulting in a serial deposition. The attractive van der Waals component of DLVO force relies on the gap between microspheres and substrate which is set by the roughness of interacting surfaces[34]. The roughness of the glass substrate is $R_a \sim 8.0$ nm (SI Figure 4(b)), which sets a lower limit on the minimum grazing distance with microspheres. The drag force at a flow rate of 10 µl/min on the very first pinned particle in the linear chain is $F_{flow} = 0.12$ nN (SI Section 2). The attractive DLVO force between microsphere and glass substrate at a gap of 8 nm in DI water is $F_{DLVO} \approx 0.09$ nN (Figure 2(d), see SI Section 3 for DLVO force calculations) which is insufficient to counter the drag by fluid flow. It is likely that the microsphere arrests due to collective effects of both DLVO force and frictional force.

The stick and slip motion exhibited by 2$^{nd}$ incoming microspheres onward in the presence of PBS medium (SI Video 3, Figure 3(a)) is a characteristic feature of frictional force[35–39]. We define contact force ($F_{con}$) as the total friction force due to attractive DLVO interaction and surface roughness acting against the drag force to arrest the microspheres. The contact force while the microsphere slips can be estimated as follows:

$$F_{con} = F_{flow} - F_{slip}$$

Where, $F_{flow} = 0.12$ nN is the drag force on a static microsphere in the presence of flow, and $F_{slip}$ is the drag force when microspheres are slipping against the pinned microspheres at speed '$V_{slip}$'. Drag force on a particle grazing near a surface can be analytically derived as $1.7 \times F_{Stokes}$, (SI Section 4) where $F_{Stokes} = 6\pi\eta rV$ is the stokes drag on a sphere moving with speed 'V', where $\eta = 8.9 \times 10^{-4}$ Pa.s is the viscosity of water, and r = 5 µm is the radius of microspheres[40–44]. The average slip velocity '$V_{slip} = 68.8 \pm 8.4$ µm/s' is estimated by taking the ratio of distance covered by microspheres while slipping through the linear chain (Figure 3(a)) and corresponding time consumed (yellow lines tracked using ImageJ in Figure 2(b)). The force required for a microsphere to attain the slip velocity can be approximated as $F_{slip} = 1.7 \times 6\pi\eta rV_{slip} = 9.8$ pN. The contact force $F_{CON} = F_{flow} - F_{slip} = 0.12$ nN $- 9.8$ pN $\approx 0.11$ nN (schematic shown in Figure 3(b)) is marginally lower than the drag due to fluid flow which is essential for the slipping to occur. The incoming microspheres graze through the linear chain and eventually deposit on the posterior end guided by fluid flow. Using fluid flow simulation discussed in SI Section 2, we estimated drag force on the linear chain as it grows. The drag force on the very first microsphere at a fluid flow rate of 10 µl/min is 120 pN and increases on an average by only 30 pN per microsphere (Figure 3(c)). Therefore, as the chain grows, the average drag force per microsphere redistributes to a lower value than drag on the very first microsphere. Hence, an increase in the length of chain doesn't adversely affect its stability.

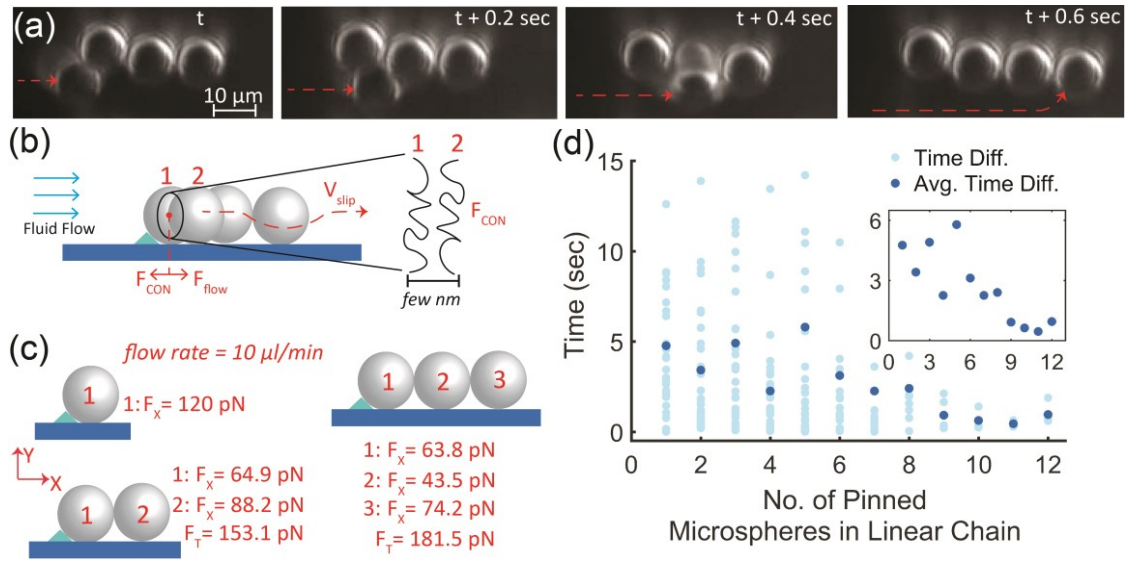

**Figure 3.** Linear chain formation in 10 μm microsphere when flown in 0.1 M PBS at a flow rate of 10 μl/min. (a) Sequence of images showing the growth of chain from 3 to 4 microspheres. (b) Microsphere grazes through the chain and deposit in the front. (c) Force redistributes along the chain such that on average drag force increases by ~30 pN per microsphere. (d) Time difference between pinning of successive microspheres evaluated over all the linear chains. Inset image shows a drop in average time difference as the length of chain grows.

The time difference between the deposition of successive microspheres is shown for all linear chains in Figure 3(d). We can see a trend of drop in average time difference as the length of chain grows (inset Figure 3(d)). The time consumed by microspheres to deposit is a function of the total number of microspheres arriving and their probability of deposition. Since there is no physical or chemical change in the system, the probability of deposition is expected to be nearly the same. A marginal increase in the drag force is anticipated as the deposition of microspheres reduces the cross-sectional area of the flow channel resulting in an increased mean fluid velocity and, consequently, the drag force.

A closer look at the tracks shown in SI Video 4 (longer version of SI Video 3) suggests a significant coupling among nearby chains. As the length of chain grows, microspheres jump from preceding chains to the forward forming chains, possibly increasing the rate of incoming microspheres. In contrast, deposition of the first few microspheres is statistical as incoming microspheres interact with a chain only if they happen to lie in their flow direction. Additionally, some chains are not entirely straight and have microspheres protruding out, which act as extra seeds for the incoming microsphere to latch on and glide their way to deposit in the front.

**Rapid Deposition in the Presence of Rough Microspheres.** Next, we used microspheres with different surface roughness to better understand the role of friction force. A mixture of 10 µm, and 4 µm microspheres of two different surface roughness values were used. We procured 4 µm microspheres of roughness 1.5 nm (Figure 4(a)) and 4.7 nm (Figure 4(b)), respectively (SI Section 5). A mixture of 0.1% w/v 4 µm microspheres of lower roughness $R_a = 1.5$ nm and 0.25% w/v 10 µm microspheres, when flown in DI medium, did not affect deposition, and only 10 µm microspheres were pinned as before (Figure 2(a), SI Section 6). In contrast, rapid deposition of microspheres is observed when the same mixture composition but with 4 µm microspheres of higher roughness ($R_a = 4.7$ nm) are flown (SI Video 5). The deposition of first few microspheres is shown in the close up movie SI Video 6. The 4 µm microspheres of lower roughness can approach 10 µm microspheres from a closer distance and are likely to have higher DLVO force as compared to rougher 4 µm microspheres. However, the occurrence of rapid deposition only in the presence of rough 4 µm microspheres suggests a dominant role of friction force in the arrest of microspheres.

Initially, microspheres are flown at the rate of 10 µl/min pinning microspheres of size 10 µm in the downslope region, whereas microspheres of size 4 µm pass through. The pinned microspheres then act as seeds for rapid deposition at a higher flow rate of 70 µl/min. The rapid deposition proceeds by pinning of 4 µm rough microspheres on already pinned 10 µm microspheres, which further advances when the incoming 10 µm microspheres deposit by latching onto 4 µm microspheres. Interestingly the rapid deposition of 4 µm and 10 µm microspheres does not occur below a flow rate of 70 µl/min. The deposition in the multi-microspheres (10 µm + 4 µm in DI water) system grows in 3 dimensions (Figure 4(c)) as opposed to the linear chains (10 µm in PBS) previously discussed (Figure 2(b)).

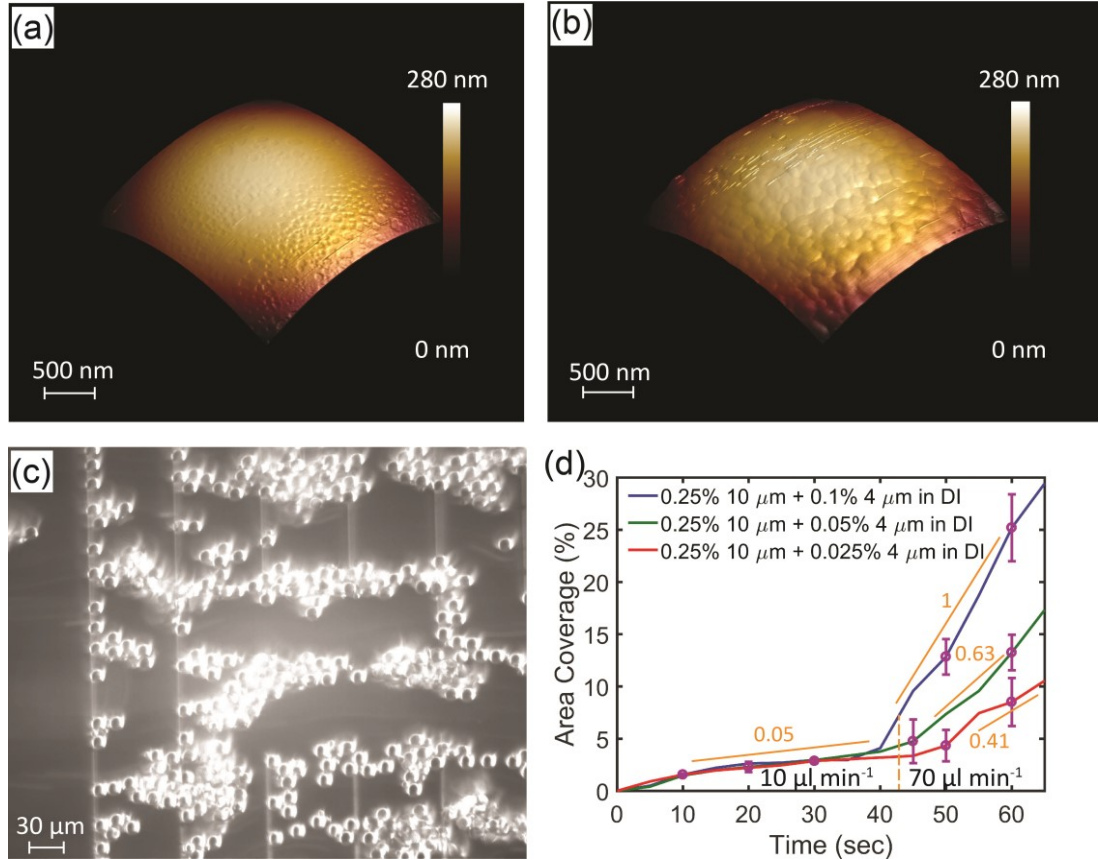

**Figure 4.** Deposition of mixed microspheres (4 μm + 10 μm) over wedge microstructures. (a) AFM of 4 μm smoother microspheres of roughness $R_a = 1.5$ nm. (b) AFM of 4 μm rough microspheres ($R_a = 4.7$ nm). (c) Rapid deposition of microspheres (4 μm + 10 μm) at a high flow rate of 70 μl min$^{-1}$. (d) The rate of Area Coverage jumps up to 20-fold as the flow rate is increased from 10 μl/min to 70 μl/min.

The rate of single pinning events of 10 μm microspheres decreases by a factor of 2 from 0.1 to 0.05 as the concentration of microspheres is halved from 0.5 % w/v (inset Figure 2(c)) to 0.25 % w/v (Figure 4(d)). Accordingly, we represent the area coverage (AC) in time 't' as:

$$AC = \delta \rho t$$

Where 'δ' is the sticking probability and 'ρ' is the volume fraction of 10 μm microspheres. The magnitude of sticking probability (δ) is a function of DLVO interaction, contact force and flow rate. If the flow rate is kept constant, then the sticking probability relies only on material properties such as salt concentration and roughness. For instance, the slope of AC vs time while keeping the concentration (ρ) fixed increases from 0.1 in DI water to 0.37 in PBS medium (Figure 2(c)). This increase is attributed to the reduction in the Debye length (PBS medium leads to lower electrostatic repulsion) resulting in an increased sticking probability. In the case of a multi-microspheres system (10 μm + 4 μm) in DI water, the deposition rates are even higher, where the area coverage rate increases up to 20 folds as the flow rate is increased to 70 µl/min (Figure 4(d)). The AC was calculated directly by thresholding images using ImageJ to construct the outer border of deposited microspheres. The rate of AC scales sub-linearly with the concentration of 4 µm microspheres, and appears to follow a power law with an exponent (2/3) as shown in Figure 4(d):

$$AC = \delta \sigma^{2/3} t$$

Where 'δ' is the sticking probability and 'σ' is the volume fraction of 4 µm microspheres. As the concentration of 4 µm microspheres is successively halved, the slope decreases to $(0.5)^{2/3} = 0.63$, $(0.25)^{2/3} = 0.40$, respectively (Figure 4(d)). The exponent (2/3) can be understood by approximating the entire deposited mass to be circular and of a radius 'R', such that the volume and area coverage of deposited mass is:

$$VOL \propto R^3 \text{ and } AC \propto R^2$$

By equating the above two equations, area coverage scales as:

$$AC = \delta \sigma^{2/3} t \propto R^2 \propto VOL^{2/3}$$

Accordingly, the volume of deposited mass is directly proportional to the concentration of 4 µm microspheres $VOL \propto \sigma$. The deposited volume (VOL) can be assumed to be made up of only 10 µm microspheres, as the volume of a single 10 µm microspheres is more than 15 times as that of a 4 µm microspheres. Therefore, on average, each 4 µm microsphere facilitates the deposition of one more 10 µm microspheres. The multi-microspheres system is unstable in 0.1 M PBS and formed clumps, rendering the deposition process erratic (SI Video 7).

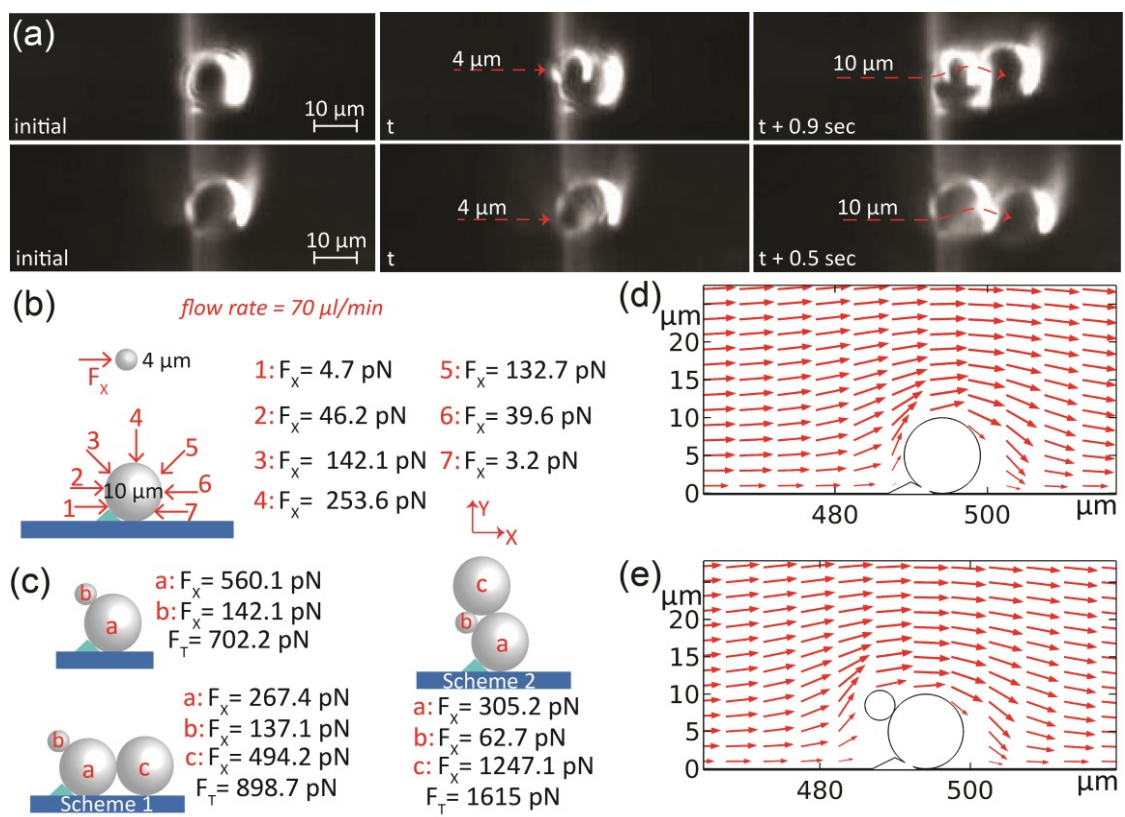

**Figure 5.** Hydrodynamic drag on 4 μm microspheres pinned over 10 μm microspheres at a flow rate of 70 μl/min. (a) Image series shows subsequent deposition of 4 μm microsphere and 10 μm microsphere on already pinned 10 μm microsphere. (b) The hydrodynamic drag force on 4 μm microsphere at different positions. Surprisingly, positions around 3 facing maximum hydrodynamic drag are the preferred position of pinning as seen from the videos. (c) Increase in drag force due to deposition of incoming 10 μm microsphere on 10 μm and 4 μm microsphere system. Microspheres face higher drag in Scheme 2 and glide to Scheme 1 configuration facing much lower drag. (d) Flow field over 10 μm microspheres. (e) Flow field over 10 μm and 4 μm microsphere system.

To understand the deposition process in the multi-microsphere system, we closely examined the initial few deposition events. At an initial flow rate of 10 μl/min, microspheres of size 10 μm pins on the microstructures. The rapid deposition then commences at a flow rate of 70 μl/min (SI Video 5), when the 4 μm microspheres pin over existing 10 μm microspheres. As seen in SI Video 6 and Figure 5(a), the preferred pinning site of 4 μm microspheres are position 3 and 4 (Figure 5(b)). The drag force, as estimated from hydrodynamic simulation (SI Section 2) is maximum at position 3 (~30° − 60° from horizontal) and position 4 (90° from horizontal) which is undesirable for pinning (Figure 5(b)). However, preference for position 3 and 4 suggests that the pinning of 4 μm microspheres is dominated by the initial contact. Further, for a stable deposition, the contact force between rough 4 μm and 10 μm microsphere should be at least ~140 pN (Figure 5(c)). The deposition process further proceeds with the 10 μm microsphere gliding its way towards the posterior region (Scheme 1 in Figure 5(c)) after initial contact with 4 μm microspheres (SI Video 6). The drag force on 10 μm microspheres near the top is ~1.25 nN (Scheme 2, Figure 5(d)) and drops to ~0.49 nN (Scheme 1, Figure 5(d)) in the posterior region where microsphere eventually settles. Even though the flow field always pushes the particles in the posterior region (Figure 5(d), Figure 5(e)), 4 μm microspheres pin at the point of contact, whereas 10 μm microspheres follow the flow direction and deposit on the posterior side.

**Components of Contact Force.**

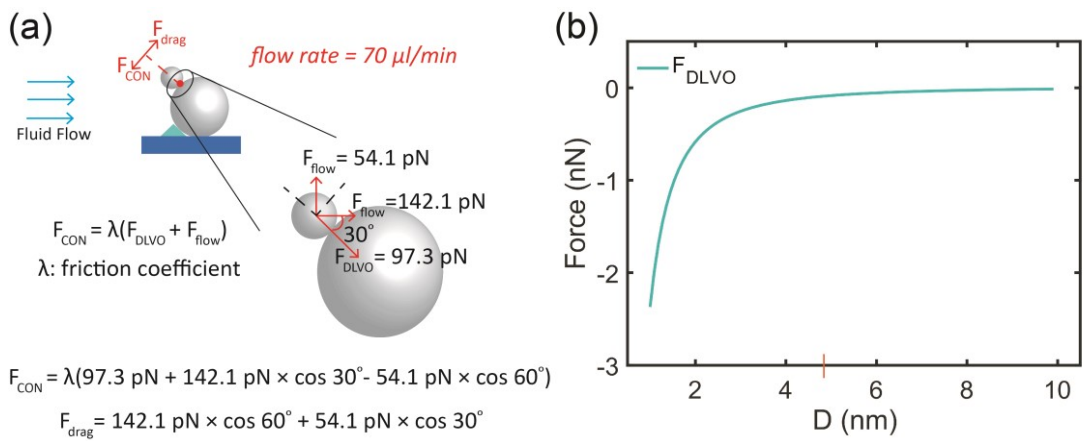

**Figure 6.** Contact force between 4 μm and 10 μm microspheres in DI water. (a) Free body diagram displaying components of contact force on a 4 μm microsphere pinned over a 10 μm microsphere. (b) DLVO force between microspheres of size 4 μm and 10 μm respectively.

The rapid deposition of mixed microspheres (4 μm + 10 μm) over microstructures is a two-step process. Initially, 10 μm microspheres pin on the microstructure at a flow rate of 10 μl/min which is followed by large scale accumulation at a higher flow rate of 70 μl/min. The rapid deposition proceeds with the pinning of a 4 μm microspheres on already pinned 10 μm microsphere (Figure 5(a)). For stable pinning of a 4 μm microsphere, the contact force ($F_{con}$) should balance the drag force ($F_{drag}$) as shown in Figure 6(a). The contact force can be expressed as the net friction force due to attractive DLVO interaction and drag generated by fluid flow:

$$F_{con} = \lambda(F_{DLVO} + F_{flow})$$

Where 'λ' is the friction coefficient. The minimum gap at which 4 μm microspheres can graze over 10 μm microspheres is limited by its roughness ~4.7 nm. The attractive DLVO force between 4 μm and 10 μm microspheres at a gap of 4.7 nm is 97.3 pN (Figure 6(b)). Similarly, the drag force generated by the fluid in parallel and perpendicular direction of the flow is numerically estimated as 142.1 pN and 54.1 pN, respectively (SI Section 2). The net friction force calculated by adding the component of DLVO and drag force perpendicular to the line of contact (Figure 6(a)) is:

$$F_{con} = \lambda(97.3 \text{ pN} + 142.1 \text{ pN} \times \cos 30° - 54.1 \text{ pN} \times \cos 60°) = \lambda \times 193.31 \text{ pN}$$

The drag force opposite to the direction of contact force is estimated by adding flow components parallel to the line of contact:

$$F_{drag} = 142.1 \text{ pN} \times \cos 60° + 54.1 \text{ pN} \times \cos 30° = 117.9 \text{ pN}$$

Accordingly, the minimum friction coefficient needed to balance the drag force is $\lambda = F_{drag}/F_{con} = 0.6$. The estimated friction coefficient '$\lambda$' is of the similar order as reported for the roughness in the range of '~10 nm'[8,45]. The DLVO force for the 4 μm microsphere of lower roughness $R_a = 1.5$ nm while grazing onto 10 μm microsphere i.e. at the gap of 1.5 nm is 993 pN. The total contact force perpendicular to the line of contact would be 1089 pN which is significantly higher than the former case. However, if the friction coefficient drops to say $\lambda = 0.1$ owing to lower roughness than the contact force $F_{con} = \lambda \times 1089 \text{ pN} = 108.9$ pN won't be able to balance the drag force ($F_{drag} = 117.9$ pN) due to fluid flow. The friction coefficient is proportional to roughness ($R_a$) but requires specialised tribometric measurements for accurate quantification[46–48]. Nevertheless, the experimental evidence suggests robust pinning of 4 μm microspheres of higher roughness leading to rapid deposition.

**CONCLUSION**

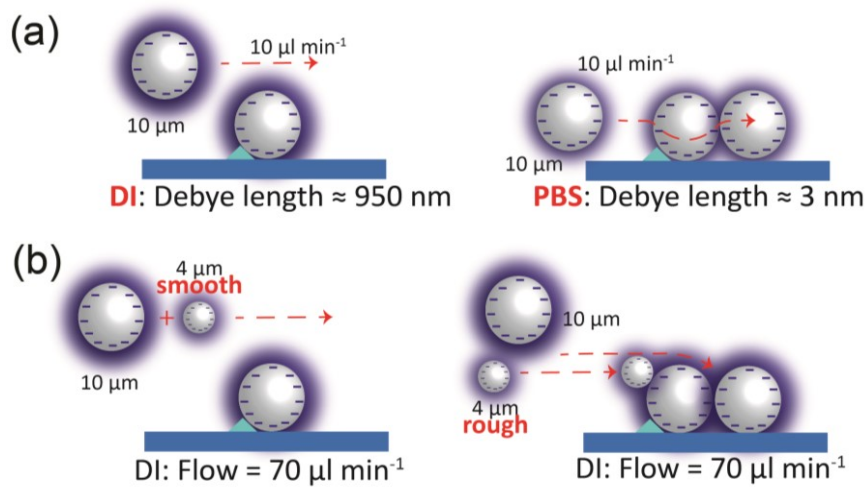

**Figure 7.** Infographic showing route towards deposition process. (a) Screening of the electric double layer repulsion by use of PBS enhances the probability of contact between microspheres, thereby assisting the formation of linear chains at a flow rate of 10 µl/min. (b) Rougher 4 µm microspheres at high flow rate of 70 µl/min pin onto 10 µm microspheres initiating the rapid deposition of incoming 10 µm microsphere.

To summarise, the deposition process initiates by pinning of 10 µm microspheres on the downslope region when flown (10 µl/min) over wedge-shaped symmetric microstructures of height 1.5 µm and width 6 µm. The incoming microspheres then interact with the pinned microspheres through attractive DLVO and roughness mediated friction force. The electrostatic component of DLVO force can be decreased by introducing salt solution such as PBS buffer (0.1 M) which leads to the formation of linear chains (Figure 7(a), Figure 3(a)). Alternatively, in DI water and at higher flow rates (70 µl/min), the presence of 4 µm microspheres of roughness $R_a$ = 4.7 nm acts as a bridge between incoming 10 µm microspheres and already pinned 10 µm microspheres (Figure 7(b), Figure 5(a)) resulting in the rapid deposition. Interestingly, a slight drop in the roughness of 4 µm microspheres from 4.7 nm to 1.5 nm doesn't lead to rapid deposition. These findings put bounds on the required ionic strength and roughness, which can lead to rapid colloidal deposition in microfluidic channels.

## ASSOCIATED CONTENT

**Supporting Information**

The supporting information is available free of charge at:

Pinning of 10 µm microspheres in DI water medium; Hydrodynamic equations for COMSOL simulation; DLVO force calculation; Analytical expression of drag force near a surface; Materials and methods; Mixture of 10 µm microspheres and microspheres of lower roughness ($R_a$ = 1.5 nm); Description of Supplementary Videos. SI Video 1 Pinning of 10 µm microspheres on the downslope at a flow rate of 10 µl/min; SI Video 2 Pinning of 10 µm microspheres on both the upslope and downslope at a flow rate of 0.1 µl/min; SI Video 3 Microsphere of Size 10 µm in PBS 0.1 M forms linear chain when passed at a flow rate of 10 µl/min; SI Video 4 Longer version of SI Video 3 to show coupling in nearby chains; SI Video 5 Microsphere of size 10 µm and 4 µm (mixed) in DI water deposits rapidly when passed at an initial flow rate of 10 µl/min and later increased to 70 µl/min; SI Video 6 Deposition of first few microspheres in the case of rapid deposition as shown in SI Video 5; SI Video 7 Microsphere of size 10 µm and 4 µm (mixed) in PBS 0.1 M aggregate to form clumps.

**AUTHOR INFORMATION**

**Corresponding Authors**

Praneet Prakash – *Centre for Nanoscience and Engineering, Indian Institute of Science, Bangalore, 560012, India. Current Affiliation - Department of Applied Mathematics and Theoretical Physics, University of Cambridge, Wilberforce Road, Cambridge, CB3 0WA, United Kingdom.* orcid.org/0000-0002-5282-0807;
Email: pp467@cam.ac.uk

Manoj Varma – *Centre for Nanoscience and Engineering, Robert Bosch Centre for Cyber Physical Systems, Indian Institute of Science, Bangalore, 560012, India.*
Email: mvarma@iisc.ac.in

**Authors**

A. Z. Abdulla – *Department of Physics, Indian Institute of Science, Bangalore, 560012, India. Current Affiliation – Laboratoire de Biologie et Modelisation de la Cellule, ENS de Lyon, 69364 Lyon Cedex 07, France.*

**ACKNOWLEDGEMENTS**

P. P. acknowledges financial support from the INSPIRE fellowship scheme of the Department of Science and Technology, Govt. of India. We thank the anonymous reviewers for their insightful comments which has significantly improved our manuscript.

**REFERENCES**

(1) Dressaire, E.; Sauret, A. Clogging of Microfluidic Systems. *Soft Matter* **2017**, 13, 37–48.

(2) Radhakrishnan, A. N.; Marques, M. P. C.; Davies, M. J.; O'Sullivan, B.; Bracewell, D. G.; Szita, N. Flocculation on a Chip: A Novel Screening Approach to Determine Floc Growth Rates and Select Flocculating Agents. *Lab Chip* **2018**, *18* (4), 585–594.

(3) Donev, A.; Torquato, S.; Stillinger, F. H.; Connelly, R. Jamming in Hard Sphere and Disk Packings. *J. Appl. Phys.* **2004**, *95* (3), 989–999.

(4) Palacci, J.; Sacanna, S.; Steinberg, A. P.; Pine, D. J.; Chaikin, P. M. Living Crystals of Light-Activated Colloidal Surfers. *Science (80 ).* **2013**, *339* (6122), 936–940.

(5) Singh, R.; Adhikari, R. Universal Hydrodynamic Mechanisms for Crystallization in Active Colloidal Suspensions. *Phys. Rev. Lett.* **2016**, *117* (22), 228002.

(6) Wang, G. R.; Yang, F.; Zhao, W. There Can Be Turbulence in Microfluidics at Low Reynolds Number. *Lab Chip* **2014**, *14* (8), 1452–1458.

(7) Mohtaschemi, M.; Puisto, A.; Illa, X.; Alava, M. J. Rheology Dynamics of

Aggregating Colloidal Suspensions. *Soft Matter* **2014**, *10* (17), 2971–2981.

(8)  Hsu, C. P.; Ramakrishna, S. N.; Zanini, M.; Spencer, N. D.; Isa, L. Roughness-Dependent Tribology Effects on Discontinuous Shear Thickening. *Proc. Natl. Acad. Sci. U. S. A.* **2018**, *115* (20), 5117–5122.

(9)  Bacchin, P.; Derekx, Q.; Veyret, D.; Glucina, K.; Moulin, P. Clogging of Microporous Channels Networks: Role of Connectivity and Tortuosity. *Microfluid. Nanofluidics* **2014**, *17* (1), 85–96.

(10) Sendekie, Z. B.; Bacchin, P. Colloidal Jamming Dynamics in Microchannel Bottlenecks. *Langmuir* **2016**, *32* (6), 1478–1488.

(11) Robert De Saint Vincent, M.; Abkarian, M.; Tabuteau, H. Dynamics of Colloid Accumulation under Flow over Porous Obstacles. *Soft Matter* **2016**, *12* (4), 1041–1050.

(12) Marin, A.; Lhuissier, H.; Rossi, M.; Kähler, C. J. Clogging in Constricted Suspension Flows. *Phys. Rev. E* **2018**, *97* (2), 021102.

(13) Souzy, M.; Zuriguel, I.; Marin, A. Transition from Clogging to Continuous Flow in Constricted Particle Suspensions. *Phys. Rev. E* **2020**, *101* (6), 060901.

(14) Sauret, A.; Somszor, K.; Villermaux, E.; Dressaire, E. Growth of Clogs in Parallel Microchannels. In *Physical Review Fluids*; American Physical Society, 2018; 3, 104301.

(15) Yodh, J. S.; Spandan, V.; Mahadevan, L. Suspension Jams in a Leaky Microfluidic Channel. *Phys. Rev. Lett.* **2020**, *125* (4), 044501.

(16) Reichhardt, C.; Reichhardt, C. J. O. Controlled Fluidization, Mobility, and Clogging in Obstacle Arrays Using Periodic Perturbations. *Phys. Rev. Lett.* **2018**, *121* (6), 068001.

(17) Zuriguel, I.; Parisi, D. R.; Hidalgo, R. C.; Lozano, C.; Janda, A.; Gago, P. A.; Peralta, J. P.; Ferrer, L. M.; Pugnaloni, L. A.; Clément, E.; et al. Clogging Transition of Many-Particle Systems Flowing through Bottlenecks. *Sci. Rep.* **2014**, *4* (1), 1–8.

(18) Sauret, A.; Barney, E. C.; Perro, A.; Villermaux, E.; Stone, H. A.; Dressaire, E. Clogging by Sieving in Microchannels: Application to the Detection of Contaminants in Colloidal Suspensions. *Appl. Phys. Lett.* **2014**, *105* (7), 074101.

(19) Wyss, H. M.; Blair, D. L.; Morris, J. F.; Stone, H. A.; Weitz, D. A. Mechanism for Clogging of Microchannels. *Phys. Rev. E* **2006**, *74* (6), 061402.

(20) Agbangla, G. C.; Bacchin, P.; Climent, E. Collective Dynamics of Flowing Colloids during Pore Clogging. *Soft Matter* **2014**, *10* (33), 6303–6315.

(21) Ortiz, C. P.; Daniels, K. E.; Riehn, R. Nonlinear Elasticity of Microsphere Heaps. *Phys. Rev. E* **2014**, *90* (2), 022304.

(22) Henry, C.; Minier, J. P.; Lefèvre, G. Towards a Description of Particulate Fouling: From Single Particle Deposition to Clogging. *Advances in Colloid and Interface Science* **2012**, 185-186, 34–76.

(23) Bizmark, N.; Schneider, J.; Priestley, R. D.; Datta, S. S. Multiscale Dynamics of Colloidal Deposition and Erosion in Porous Media. *Sci. Adv.* **2020**, *6* (46), 2530–2543.

(24) Ramachandran, V.; Fogler, H. S. Multilayer Deposition of Stable Colloidal Particles during Flow within Cylindrical Pores. *Langmuir* **1998**, *14* (16), 4435–4444.

(25) Cejas, C. M.; Monti, F.; Truchet, M.; Burnouf, J. P.; Tabeling, P. Particle Deposition Kinetics of Colloidal Suspensions in Microchannels at High Ionic Strength. *Langmuir* **2017**, *33* (26), 6471–6480.

(26) Prakash, P.; Varma, M. Trapping/Pinning of Colloidal Microspheres over Glass Substrate Using Surface Features. *Sci. Rep.* **2017**, *7* (1), 15754.

(27) Clavauda, C.; Be-ruta, A.; Metzgera, B.; Forterrea, Y. Revealing the Frictional Transition in Shear-Thickening Suspensions. *Proc. Natl. Acad. Sci. U. S. A.* **2017**, *114* (20), 5147–5152.

(28) Burke, A. P.; Farb, A.; Malcom, G. T.; Liang, Y.; Smialek, J.; Virmani, R. Coronary Risk Factors and Plaque Morphology in Men with Coronary Disease Who Died Suddenly. *N. Engl. J. Med.* **1997**, *336* (18), 1276–1282.

(29) Libby, P.; Ridker, P. M.; Hansson, G. K. Progress and Challenges in Translating the Biology of Atherosclerosis. *Nature* **2011**, 473, 317–325.

(30) Kojima, Y.; Volkmer, J. P.; McKenna, K.; Civelek, M.; Lusis, A. J.; Miller, C. L.; Direnzo, D.; Nanda, V.; Ye, J.; Connolly, A. J.; et al. CD47-Blocking Antibodies Restore Phagocytosis and Prevent Atherosclerosis. *Nature* **2016**, *536* (7614), 86–90.

(31) Hajhosseiny, R.; Bahaei, T. S.; Prieto, C.; Botnar, R. M. Molecular and Nonmolecular Magnetic Resonance Coronary and Carotid Imaging. *Arterioscler. Thromb. Vasc. Biol.* **2019**, *39* (4), 569–582.

(32) Prakash, P.; Pahal, S.; Varma, M. Fluorescence Recovery after Photobleaching in Ultrathin Polymer Films. *Macromol. Chem. Phys.* **2018**, *219* (7), 1700543.

(33) Israelachvili, J. N.; Intermolecular and Surface Forces. *Elsevier Inc.* **2011**, 3rd Edition.

(34) Bhattacharjee, S.; Ko, C.-H.; Elimelech, M. DLVO Interaction between Rough Surfaces. *Langmuir* **1998**, 14(12), 3365-3375.

(35) Kligerman, Y.; Varenberg, M. Elimination of Stick-Slip Motion in Sliding of Split or Rough Surface. *Tribol. Lett.* **2014**, *53* (2), 395–399.

(36) Lee, D. W.; Banquy, X.; Israelachvili, J. N. Stick-Slip Friction and Wear of Articular Joints. *Proc. Natl. Acad. Sci. U. S. A.* **2013**, *110* (7), E567–E574.

(37) Thompson, P. A.; Robbins, M. O. Origin of Stick-Slip Motion in Boundary Lubrication. *Science (80-. ).* **1990**, *250* (4982), 792–794.

(38) Bengisu, M. T.; Akay, A. Stick–Slip Oscillations: Dynamics of Friction and Surface Roughness. *J. Acoust. Soc. Am.* **1999**, *105* (1), 194–205.

(39) Chen, Z.; Khajeh, A.; Martini, A.; Kim, S. H. Chemical and Physical Origins of Friction on Surfaces with Atomic Steps. *Sci. Adv.* **2019**, *5* (8), eaaw0513.

(40) Mulvaney, S. P.; Cole, C. L.; Kniller, M. D.; Malito, M.; Tamanaha, C. R.; Rife, J. C.; Stanton, M. W.; Whitman, L. J. Rapid, Femtomolar Bioassays in Complex Matrices Combining Microfluidics and Magnetoelectronics. *Biosens. Bioelectron.* **2007**, *23* (2), 191–200.

(41) Gijs, M. A. M.; Lacharme, F.; Lehmann, U. Microfluidic Applications of Magnetic Particles for Biological Analysis and Catalysis. *Chem. Rev.* **2010**, *110* (3), 1518–1563.

(42) White F. M., Fluid Mechanics. *Mc Graw Hill* **2016,** 8th Edition.

(43) Prakash, P.; Abdulla, A. Z.; Singh, V.; Varma, M. Tuning the Torque-Speed Characteristics of the Bacterial Flagellar Motor to Enhance Swimming Speed. *Phys. Rev. E* **2019**, *100* (6), 062609.

(44) Prakash, P.; Abdulla, A. Z.; Singh, V.; Varma, M. Swimming Statistics of Cargo-

Loaded Single Bacteria. *Soft Matter* **2020**, *16* (41), 9499–9505.

(45) Zhang, Y.; Sundararajan, S. The Effect of Autocorrelation Length on the Real Area of Contact and Friction Behavior of Rough Surfaces. *J. Appl. Phys.* **2005**, *97* (10), 103526.

(46) Berman, A. D.; Ducker, W. A.; Israelachvili, J. N. Origin and Characterization of Different Stick-Slip Friction Mechanisms. *Langmuir* **1996**, *12* (19), 4559–4562.

(47) Fernandez, N.; Cayer-Barrioz, J.; Isa, L.; Spencer, N. D. Direct, Robust Technique for the Measurement of Friction between Microspheres. *Langmuir* **2015**, *31* (32), 8809–8817.

(48) Zhang, Y.; Sundararajan, S. Adhesion and Friction Studies of Silicon Surfaces Processed Using a Microparticle-Based Method. *Tribol. Lett.* **2006**, *23* (1), 1–5.

**TOC GRAPHIC**

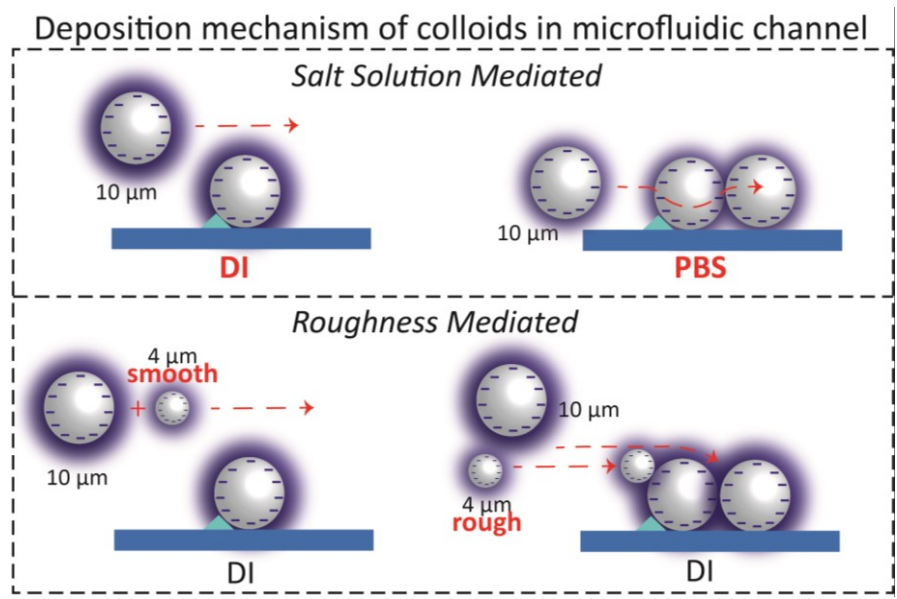